\documentclass[aps,prfluids,showpacs]{revtex4-1}
\usepackage{amsfonts}
\usepackage{booktabs}
\usepackage{amsmath}
\usepackage{amsthm}
\usepackage{graphics}
\usepackage{graphicx}
\usepackage{subfigure}
\usepackage{amssymb}
\usepackage{graphics}
\usepackage{graphicx}
\usepackage{epstopdf}
\usepackage{color}
\usepackage{subfigure}
\usepackage{pgfplots,tikz}
\usepackage{url}
\usepackage{float}
\usepackage{soul}
\def\x{{\bf x}}

\newcommand\dertt[1]{ \frac{\partial{ #1}}{\partial t} }

\def\Vp{V_{\rm p}}
\def\Mp{M_{\rm p}}
\def\Rp{a_{\rm p}}

\begin{document}

\title{Active and finite-size particles in decaying quantum turbulence at low temperature}
\author{Umberto Giuriato}
\affiliation{
Universit\'e C\^ote d'Azur, Observatoire de la C\^ote d'Azur, CNRS, Laboratoire Lagrange, Nice, France}
\author{Giorgio Krstulovic}
\affiliation{
Universit\'e C\^ote d'Azur, Observatoire de la C\^ote d'Azur, CNRS, Laboratoire Lagrange, Nice, France}

\begin{abstract}
The evolution of a turbulent tangle of quantum vortices in presence of finite-size active particles is studied by means of numerical simulations of the Gross-Pitaevskii equation. Particles are modeled as potentials depleting the superfluid and described with classical degrees of freedom following a Newtonian dynamics. It is shown that particles do not modify the building-up and the decay of the superfluid Kolmogorov turbulent regime. It is observed that almost the totality of particles remains trapped inside quantum vortices, although they are occasionally detached and recaptured. The statistics of this process are presented and discussed.
The particle Lagrangian dynamics is also studied. At large time scales, the velocity spectrum of particles is reminiscent of a classical Lagrangian turbulent behavior. At time-scales faster than the turnover time associated to the mean inter-vortex distance, the particle motion is dominated by oscillations due to Magnus effect. For light particles a non-classical scaling of the spectrum arises. The particle velocity and acceleration probability distribution functions are then studied. The decorrelation time of the particle acceleration is found to be shorter than in classical fluids, and related to the Magnus force experienced by the trapped particles.
\end{abstract}
\maketitle

\section{Introduction}

When a fluid is stirred, energy is injected into the system exciting structures at different scales. In particular, in three dimensional classical flows, the energy supplied at large scales is transferred towards small scales in a cascade process. Eventually, it reaches the smallest scales of the system, where dissipation acts efficiently. In presence of a very large separation between the injection and dissipation scale, this cascade scenario proposed by Richardson, leads to a fully developed turbulent state that can be described by the Kolmogorov phenomenology \cite{frisch1995turbulence}. Kolmogorov turbulence is expected to be universal, and it is in fact commonly observed in nature, industrial applications and in more exotics flows such as superfluids.

A superfluid is a peculiar flow, whose origin is a consequence of quantum mechanics. At finite temperature, a superfluid is considered to be a mixture of two components: the normal fluid, that can be described by the Navier-Stokes equations, and the superfluid component with zero viscosity \cite{donnelly1991quantized}. At very low temperatures, the normal component can be neglected and the fluid becomes completely inviscid. As a consequence, an object moving at low velocities does not experience any drag from the fluid. However, when the object exceeds a critical velocity, quantum vortices are nucleated \cite{FrischPomeauRicaVortex,ActiveWiniecki}. Quantum vortices (or superfluid vortices) are the most fundamental hydrodynamical excitations of a superfluid. They are topological defects (and nodal lines) of the macroscopic wave function describing the system, and as a consequence their circulation is quantized. In superfluid helium, the core size of quantum vortices is of the order of 1\AA.
Despite the lack of viscosity, quantum vortices can reconnect and change their topology (see for instance \cite{KoplikLevine,BewleyReconnectionExp,ReconnectionGiorgio,ReconnectionLuca}), unlike classical (prefect) fluids.

When energy is injected in a low-temperature superfluid at scales much larger than the mean inter-vortex distance $\ell$, a classical Kolmogorov regime is expected. Such a behavior has been observed numerically \cite{Noretal,BagalleyLaurieBarenghiK41andvortAnalysis,ShuklaKrstulovicEffVisco} and experimentally \cite{Maurer_1998,SalotTurbu}. Indeed, at such scales the quantum nature of vortices is not important and the superfluid behaves like a classical fluid. At the scales of the order of $\ell$ and smaller, the isolated nature of quantized vortices become relevant. The system keeps transferring energy towards small scales but through different non-classical mechanisms \cite{VinenKW}. An example of such mechanisms is the turbulent Kelvin wave cascade. Kelvin waves are helical oscillations propagating along quantum vortices and the energy can be carried toward small scales thanks to non-linear wave interactions. This energy cascade has been successfully described in the framework of weak-wave turbulence theory \cite{LvovNazarenkoKW,LaurieKWPRB}. The resulting theoretical prediction have been observed numerically in vortex-filament and Gross-Pitaevskii numerical simulations \cite{KrstulovicKW, BaggaleyLaurieKW,VilloisTangle}. 

Flow visualization is certainly a fundamental issue in every fluid dynamics experiment. Among the techniques which have been developed to sample a fluid, particle image velocimetry (PIV) and particle tracking velocimetry (PTV) are two of the most common methods \cite{ParticlesToschi}. The use of particles as probes has been also adapted to the study of cryogenic flows, in particular in superfluid helium $^4\mathrm{He}$ experiments \cite{GuoReview}, where micrometer sized hydrogen and deuterium particles have been used.
For instance, hydrogen ice particles have been successfully employed to visualize isolated or reconnecting vortex lines \cite{bewley2006superfluid}, as well as the propagation of Kelvin waves \cite{FondaKWExp}. Moreover, the observation of power-law tails in the probability density of the particle velocity is an important difference with respect to classical turbulent states \cite{PaolettiVelStat2008,LaMantiaVelocity,LaMantiaQT}. Similar deviations from classical behaviors have been recently reported also for the acceleration statistics \cite{LaMantiaQT,LaMantiaAcceleration}. Particles in such experiments typically have a size that can rise up to several microns, which is many orders of magnitude larger than the size of the vortex core in superfluid helium. For instance, the solidified hydrogen particles produced in the experiments \cite{bewley2006superfluid,FondaKWExp} are slightly smaller than $2.7\,\mu m$, while in \cite{LaMantiaVelocity,LaMantiaQT} their size is between $5\,\mu m$ and $10\,\mu m$.
Although it has been seen that particles unveils the dynamics of quantum vortices, it is yet not clear how much they affect the dynamics of quantum turbulent flows.

Several theoretical efforts have been made in the last decade in order to clarify what is the dynamics of particles in a superfluid and how particles interact with quantum vortices. For example, the vortex-filament (VF) method can be coupled with the classical hydrodynamical equations of a sphere, allowing to study different specific problems. The interaction between one particle and one vortex has been addressed \cite{ReviewNewcastle, CloseParticleNewcastle}, as well the back-reaction of tracers in a thermal counterflow \cite{BarenghiTracersFirst,BarenghiTracers}. In the context of finite temperature superfluids, the spatial statistics of particles have been recently addressed in simulations of the Hall-Vinen-Bekarevich-Khalatnikov model \cite{NachoHVBK}.

Finally, since the work of Winiecki and Adams \cite{ActiveWiniecki}, particles described by classical degrees of freedom have been implemented self-consistently in the framework of the Gross-Pitaevskii (GP) equation \cite{ShuklaSticking,GiuriatoApproach,GiuriatoClustering,GiuriatoCrystal,griffin2019magnus}. Although the GP model is formally derived for dilute Bose-Einstein condensates, it is considered as a general tool for the study of superfluid dynamics at very low temperature. Indeed, unlike the VF method or the HVBK model, it naturally contains quantum vortices as topological defects of the order parameter. It was found analytically and confirmed numerically that the GP model can reproduce the process of trapping of large active inertial particles by straight vortex lines \cite{GiuriatoApproach}, in accordance with hydrodynamical calculations \cite{ReviewNewcastle, CloseParticleNewcastle}. In this framework, the interplay between many trapped particles and Kelvin waves has been also investigated \cite{GiuriatoCrystal}.

In the present work, we study the influence of particles on quantum turbulent flows at very low temperature by using the GP model coupled with classical particles.
In particular, we study the evolution of a free decaying superfluid turbulent vortex tangle loaded with finite-size active particles. We consider spherical particles of different masses and having a diameter up to $20$ core sizes. Such size is about $1000$ times smaller than the one of solidified particles used in superfluid helium experiments. Nevertheless, it is slightly smaller than or comparable to the mean inter-vortex distance in our simulations, similar to current experiments. We also study the different regimes of the turbulent evolution from the Lagrangian point of view. The paper is organized as follows. In Section \ref{Sec:Model} we describe the Gross-Pitaevskii model coupled with classical particles. We also review the standard properties of the model and give the basic definitions used later to analyze the flow. We also describe the numerical method used in this work. Then, in Section \ref{Sec:NumericalResuts}, we present our main results. In particular, in Section \ref{SubSec:KolmogorovAndParticles} we address whether or not the presence of particles affects the scales of the flow at which Kolmogorov turbulence takes place. Section \ref{Subsec:MotionOfParticles} is devoted to the study of the particle dynamics inside the vortex tangle, to their trapping by vortices and to their dynamics at scales larger and smaller than the inter-vortex distance. Particle velocity and acceleration statistics are then presented in Section \ref{Subsec:PartStatistics}. Finally, Section \ref{Sec:Discussion} contains our conclusions.

\section{Model for particles in a low temperature superfluid \label{Sec:Model}}
\subsection{Gross-Pitaevskii equation coupled with particles}
We describe a superfluid of volume $V$ at low-temperature by using the complex field $\psi$, which obeys the GP dynamics. We consider $N_\mathrm{p}$ particles in the system. Each particle is characterized by the position of its center of mass $\mathbf{q}_i$ and its classical momentum $\mathbf{p}_i$. 
The presence of a particle of size $\Rp$ generates a superfluid depletion in a spherical region of radius $\Rp$. This effect is reproduced by coupling the superfluid field with a strong localized potential $\Vp$, which has a fixed shape and is centered at the position $\mathbf{q}_j(t)$.

All the particles considered have the same size, as well as the same mass $M_\mathrm{p}$.  The Hamiltonian of the system is given by
\begin{equation}
H=\int\left(\frac{\hbar^2}{2m}|\nabla\psi|^2+\frac{g}{2}\left(|\psi|^2 - \frac{\mu}{g}\right)^2 + {\sum_{i=1}^{N_\mathrm{p}}V_\mathrm{p}(|\mathbf{x}-\mathbf{q}_i|)|\psi|^2} \right)\mathrm{d}\mathbf{x} + {\sum_{i=1}^{N_\mathrm{p}}\frac{\mathbf{p}_i^2}{2M_\mathrm{p}}}, 
+ {\sum_{i<j}^{N_\mathrm{p}}V_\mathrm{rep}^{ij}}.
\label{Eq:Hamiltonian}
\end{equation}
where $m$ is the mass of the bosons constituting the superfluid and $g$ is the nonlinear coupling constant between the bosons, related to the $s$-wave scattering length $a_{\rm s}$ so that $g=4 \pi  a_\mathrm{s} \hbar^2 /m$. The chemical potential is denoted by $\mu$. The particle interaction potential $V_\mathrm{rep}^{ij}$ is responsible for short-range repulsion between particles, so that they behave as hard spheres and do not overlap. A detailed discussion on the inclusion of this short range repulsion and the effect on the particle collisions in the model (\ref{Eq:Hamiltonian}) can be found in \cite{ShuklaSticking}.
The equations of motion that govern the superfluid field and the particle positions are obtained varying the Hamiltonian (\ref{Eq:Hamiltonian}):
\begin{eqnarray}
i\hbar\dertt{\psi} = - \frac{\hbar^2}{2m}\nabla^2 \psi +  (g|\psi|^2 - \mu)\psi+\sum_{i=1}^{N_\mathrm{p}}\Vp(| \x -{\bf q}_i |)\psi \label{Eq:GPPfield},\\
\Mp\ddot{\bf q}_i = - \int  \Vp(| \x -{\bf q}_i|) \nabla|\psi|^2\, \mathrm{d} \x+\sum_{j\neq i}^{N_\mathrm{p}}\frac{\partial}{\partial{\bf q}_i }V_\mathrm{rep}^{ij}, 
\label{Eq:GPPparticles}
\end{eqnarray}
This model has been successfully used to study vortex nucleation \cite{ActiveWiniecki}, trapping of particles by quantum vortices \cite{GiuriatoApproach} and the interaction between particles trapped inside quantum vortices and Kelvin waves \cite{GiuriatoCrystal}. We denote by GP the Gross-Pitaevskii model without particles and by GP-P the full coupled system (\ref{Eq:GPPfield}-\ref{Eq:GPPparticles}).

In the case where particles are absent, the chemical potential $\mu$ fixes the value of the ground state of the system $\psi_\infty=\sqrt{{\rho_\infty}/{m}}=\sqrt{\mu/g}$. Large wavelength perturbations around this state are sound waves that propagate with the speed of sound $c=\sqrt{g\rho_\infty/m^2}$, while they become dispersive at length scales smaller than the healing length $\xi=\sqrt{\hbar^2/2g\rho_\infty }$. 

The GP model describes a superfluid with zero viscosity. Using the  Madelung transformation $\psi(\x)=\sqrt{{\rho(\x)}/{m}}\,e^{i\frac{m}{\hbar}\phi(\x)}$ the GP equation \eqref{Eq:GPPfield} is mapped into the continuity and Bernoulli equations of a superfluid of density $\rho$ and velocity $\mathbf{v}_\mathrm{s}=\nabla\phi$. A superfluid flow is potential, but the phase is not defined at the nodal lines of $\psi(\x)$. Therefore the vorticity is concentrated along these filaments, which are the topological defects usually called quantum vortices. The effective size of the quantum vortex core coincides with the healing length $\xi$, and the contour integral of the superfluid velocity around a single vortex filament is the Feynman-Onsager quantum of circulation $\kappa = h/m= 2\pi \sqrt{2} c\xi$. 

Using the Madelung transformation and the Helmholtz decomposition, the kinetic term of the superfluid energy density is decomposed into incompressible, compressible and quantum energy \cite{Noretal}: 
\begin{equation}
E_\mathrm{kin}^\mathrm{GP}=\frac{\hbar^2}{2mV}\int |\nabla\psi|^2\,\mathrm{d}\mathbf{x} = E_\mathrm{kin}^\mathrm{I} + E_\mathrm{kin}^\mathrm{C} + E^\mathrm{Q} = \frac{1}{2V}\int \left( \nabla(\sqrt{\rho}\mathbf{v}_\mathrm{s})^\mathrm{I} + \nabla(\sqrt{\rho}\mathbf{v}_\mathrm{s})^\mathrm{C} + \frac{\kappa}{2\pi}\nabla\sqrt{\rho} \right)\,\mathrm{d}\mathbf{x},
\label{Eq:energy}
\end{equation}
where $(\sqrt{\rho}\mathbf{v}_\mathrm{s})^\mathrm{I} = \mathcal{P}_\mathrm{I}[\sqrt{\rho}\mathbf{v}_\mathrm{s}]$ and  $(\sqrt{\rho}\mathbf{v}_\mathrm{s})^\mathrm{C} = \mathbf{v}_\mathrm{s} - (\sqrt{\rho}\mathbf{v}_\mathrm{s})^\mathrm{I}$, the operator $\mathcal{P}_\mathrm{I}[\cdot]$ being the projector onto the space of divergence-free fields. The other energies of the superfluid are the internal energy $E_\mathrm{int}=(2V)^{-1}\int g(\rho/m - \mu/g)^2\,\mathrm{d}\mathbf{x}$, where the energy of the ground-state is subtracted, and the interaction energy with the particles $E_\mathrm{P}^\mathrm{GP}=V^{-1}\int\sum_i^{N_\mathrm{p}}V_\mathrm{p}(|\mathbf{x}-\mathbf{q}_i|)\rho\,\mathrm{d}\mathbf{x}$, so that the total energy is given by $E_\mathrm{tot}=E_\mathrm{kin}^\mathrm{GP}+E_\mathrm{int}+E_\mathrm{P}^\mathrm{GP}$. From these definitions follow the corresponding energy spectra defined in terms of the Fourier transform of the fields \cite{Noretal}.

\subsection{Numerical methods and parameters}
In the simulations presented in this work we solve the system (\ref{Eq:GPPfield}-\ref{Eq:GPPparticles}) in a cubic periodic box of side $L=341\xi$ with $N_\mathrm{c}=512^3$ collocation points by using a standard pseudo-spectral method. We use a $4^\mathrm{th}$ order Runge-Kutta scheme for the time-stepping and the standard $2/3$ rule for the dealiasing. In numerics, we fix $c=1$ and $\psi_\infty=1$. 

In order to produce a homogeneous and isotropic tangle of quantized vortex lines, we impose an initial Arnold-Beltrami-Childress (ABC) flow, following the procedure described in \cite{MarcABC}. In particular, we use a superposition of $k = 1\cdot 2\pi/L$ and $k = 2\cdot 2\pi/L$ basic ABC flows: $\mathbf{v}_\mathrm{ABC}=\mathbf{v}^{(1)}_\mathrm{ABC} +  \mathbf{v}^{(2)}_\mathrm{ABC}$, with 
\begin{equation}
\mathbf{v}_\mathrm{ABC}^{(k)} = [B \cos(ky) + C \sin(kz)]\hat{x} + [C \cos(kz) + A \sin(kx)]\hat{y} + [A \cos(kx) + B \sin(ky)]\hat{z}, 
\label{Eq:vABC}
\end{equation}
and the parameters $A=0.5196$, $B=0.5774$, $C=0.6351$. The basic ABC flow is a stationary (periodic) solution of the Euler equation with maximal helicity. 
The resulting wave function contains a tangle whose nodal lines follows the ABC vortex lines. The initial mean inter--vortex distance is $\ell (t=0) \sim 25\xi$. As the flow is prepared by minimizing the energy,  most of the energy of the system is in the incompressible part of the energy and resulting form the vortex configuration.

The ground state for the particles consists in a number of particles (we use $N_\mathrm{p}=200$ and $N_\mathrm{p}=80$) of the same size and mass, randomly distributed in the computational box. Particles are initially at rest. This state is prepared using the imaginary-time evolution of the equation (\ref{Eq:GPPfield}). Then, the initial condition for the simulations is obtained by multiplying the wave function associated with the ABC flow and the wave function associated to the particles ground state. An example of an initial field containing particles is displayed in Fig.\ref{Fig:3DVizFull}.d.

Because of the presence of a healing layer, the particle boundary is never sharp, independently of the functional form of the potential $V_\mathrm{p}$. The superfluid field vanishes in the region where $V_\mathrm{p}>\mu$ and at the particle boundary the fluid density passes from zero to the bulk value $\rho_\infty$ in approximately one healing length.
The potential used to model each particle is a smoothed hat-function $\Vp(r)=\frac{V_0}{2}(1-\tanh\left[\frac{r^2 -\zeta^2}{4\Delta_a^2}\right])$ where the parameters $\zeta$ and $\Delta_a$ are set to model the particle. Their values are listed in Table \ref{tab:DNS_parameters}. In particular, $\zeta$ fixes the width of the potential and it is related to the particle size, while $\Delta_a$ controls the steepness of the smoothed hat-function. The latter needs to be adjusted in order to avoid Gibbs effect in the Fourier transform of $V_\mathrm{p}$. Since the particle boundaries are not sharp, the effective particle radius is defined as $\Rp=(3M_0/4\pi\rho_\infty)^\frac{1}{3}$, where $M_0=\rho_\infty L^3(1-\int |\psi_\mathrm{p}|^2\,\mathrm{d}\mathbf{x}/\int |\psi_\infty|^2\,\mathrm{d}\mathbf{x})$ is the fluid mass displaced by the particle and $\psi_\mathrm{p}$ is the steady state with just one particle. Practically, given the set of numerical parameters $\zeta$ and $\Delta_a$, the state $\psi_\mathrm{p}$ is obtained numerically with imaginary time evolution and the excluded mass $M_0$ is measured directly.  Particles attract each other by a short range fluid mediated interaction \cite{ShuklaSticking,GiuriatoClustering}, thus we use the repulsive potential $V_\mathrm{rep}^{ij}=\gamma(2\Rp/|{\bf q}_i-{\bf q}_j|)^{12}$ in order to avoid an overlap between them. The functional form of $V_\mathrm{rep}^{ij}$ is inspired by the repulsive term of the Lennard-Jones potential and the pre-factor $\gamma$ is adjusted numerically so that  the inter-particle distance $2\Rp$ minimizes the sum of $V_\mathrm{rep}^{ij}$ with the fluid mediated attractive potential \cite{ShuklaSticking,GiuriatoClustering}.
We express the particle mass as $M_\mathrm{p}=\mathcal{M}M_0$ , where $M_0$ is the mass of the superfluid displaced by the particle. Namely, heavy particles have $\mathcal{M}> 1$ and light particles have $\mathcal{M}< 1$. 
In Table \ref{tab:DNS_parameters} all the parameters for the particles used in the simulations presented in this work are reported.
\begin{table}[htb]
  \caption{%
    Simulation parameters.
  }\label{tab:DNS_parameters}
  \newcommand*{\ct}[1]{\multicolumn{1}{c}{#1}}  
  \begin{ruledtabular}  
  \begin{tabular}{lccccccc}
	  Run & $N_\mathrm{p}$ & $a_\mathrm{p}$ & $\mathcal{M}$ &
	  $\quad\zeta\quad$ & $\quad\Delta_a\quad$ & $\quad V_0/\mu\quad$ & $\gamma/\mu$ \\
    \midrule                   
    I    & 0   & --   & --    & --    & --   & --    & --     \\
    II   & 200 & $4.0\xi$   & 0.125 & $1.5\xi$   & $1.2\xi$  & 20.0  & $1.4\cdot 10^{-4}$ \\ 
    III  & 200 & $4.0\xi$   & 0.25  & $1.5\xi$   & $1.2\xi$  & 20.0  & $1.4\cdot 10^{-4}$ \\ 
    IV   & 200 & $4.0\xi$   & 1.0   & $1.5\xi$   & $1.2\xi$  & 20.0  & $1.4\cdot 10^{-4}$ \\ 
    V    & 200 & $4.0\xi$   & 2.0   & $1.5\xi$   & $1.2\xi$  & 20.0  & $1.4\cdot 10^{-4}$ \\
    VI   & 80  & $10.0\xi$  & 1.0   & $8.0\xi$   & $2.0\xi$  & 20.0  & $5.8\cdot 10^{-4}$ \\
    VII  & 200 & $10.0\xi$  & 0.125 & $8.0\xi$   & $2.0\xi$  & 20.0  & $5.8\cdot 10^{-4}$ \\
    VIII & 200 & $10.0\xi$  & 0.25  & $8.0\xi$   & $2.0\xi$  & 20.0  & $5.8\cdot 10^{-4}$ \\
    IX   & 200 & $10.0\xi$  & 1.0   & $8.0\xi$   & $2.0\xi$  & 20.0  & $5.8\cdot 10^{-4}$ \\
  \end{tabular}
  \end{ruledtabular}
\end{table}
In the following, we will refer to each simulation specifying the size and the mass of the particles used.

Note that although the model (\ref{Eq:Hamiltonian}) is a minimal model for implementing particles in the GP framework, we can not add to the system an arbitrary number of particles. Indeed, since particles have a finite size, they occupy a volume at the expense of the superfluid field and packing effects could become important if the filling fraction is too high. Moreover, the potential $V_\mathrm{p}$ must be updated at each time step, which is numerically costly. Finally, note that the the evaluation of the force term (\ref{Eq:GPPparticles}) acting on particles requires to know the value of the fields at inter-mesh points. When the number of particles in the simulation is not large, the force $\mathbf{f}_i^\mathrm{GP}(\mathbf{q}_i)=-\left(V_\mathrm{p}*\nabla\rho\right)\left[\mathbf{q}_i\right]$ (\ref{Eq:GPPparticles}) can be computed with spectral accuracy using a Fourier interpolation. Such method has been used in \cite{GiuriatoApproach,GiuriatoClustering,GiuriatoCrystal}, where the particle dynamics is extremely sensitive. In this work, the use of a Fourier interpolation for each particle is numerically unaffordable, due to the large number of particles involved and the resolutions used. Instead, we use a fourth--order B--spline interpolation method, which has been shown to be highly accurate with a reduced computational cost \cite{ToschiBspline} and particularly well adapted for pseudo--spectral codes. Indeed, the use of a Fourier interpolation to evaluate the three dimensional force for $N_\mathrm{p}$ particles requires $\sim 3N_\mathrm{p}N_{\mathrm{c}}$ operations and evaluations of complex exponentials ($N_{\mathrm{c}}=512^3$ in the present work). Such cost quickly becomes too expensive at high resolutions and/or large number of particles. On the contrary,  B--Spline interpolation requires just one Fast Fourier Transform of a field per component, and an interpolation using only four neighboring grid points per dimension \cite{ToschiBspline}. Such scheme saves a factor $\sim N_\mathrm{p}$ of computational cost compared to Fourier interpolation. Note that in the previous discussion we have not taken into consideration parallelization issues, where local schemes (B--Splines) are much advantageous than global ones (Fourier transforms). Nevertheless, some issues with physical quantities at small scales arising from the B--spline interpolation are discussed in Appendix.

\section{Particles immersed in a tangle of superfluid vortices \label{Sec:NumericalResuts}}

Superfluid turbulence in the context of the GP model has been largely studied \cite{Noretal,SasaQuantumTurbulence,MarcABC,GridGiorgio,ShuklaKrstulovicEffVisco}. In general, quantum turbulence develops from an initial state with a vortex configuration where the incompressible kinetic energy is mainly contained at large scale. During the evolution, vortex lines move, interact among themselves and reconnect creating complex vortex tangles. Through this process, sound is produced and incompressible kinetic energy is irreversibly converted into quantum, internal  and compressible kinetic energy. Eventually, the compressible energy produced in the form of acoustic fluctuations starts to dominate, thermalizes and acts as thermal bath providing an effective dissipation acting vortices. As a consequence, vortices shrink and eventually disappear through mutual friction effects following Vinen's decay law \cite{vinen1957mutual,VilloisTangle}. In particular, it has been shown that the decrease of the incompressible kinetic energy behaves in a similar manner to decaying classical turbulence \cite{Noretal}. In order to make connection with decaying classical Kolmogorov turbulence, the incompressible energy dissipation or dissipation rate is usually defined in the context of GP turbulence as
\begin{equation}
\epsilon = -\frac{\mathrm{d}E_\mathrm{kin}^\mathrm{I}}{\mathrm{d}t}.
\end{equation} 
As in decaying Navier-Stokes turbulence, in GP the most turbulent stage is achieved around the time when this quantity is maximal. About this time, the classical picture holds and the incompressible energy spectrum satisfies the Kolmogorov prediction 
$$
E_\mathrm{kin}^\mathrm{I}=C \epsilon^{2/3}k^{-5/3},
$$
where $C$ is the Kolmogorov constant which value has been found to be close to $1$ in GP turbulence \cite{GridGiorgio,MarcABC,ShuklaKrstulovicEffVisco}.

\begin{figure}[h!]
\includegraphics[width=.99\linewidth]{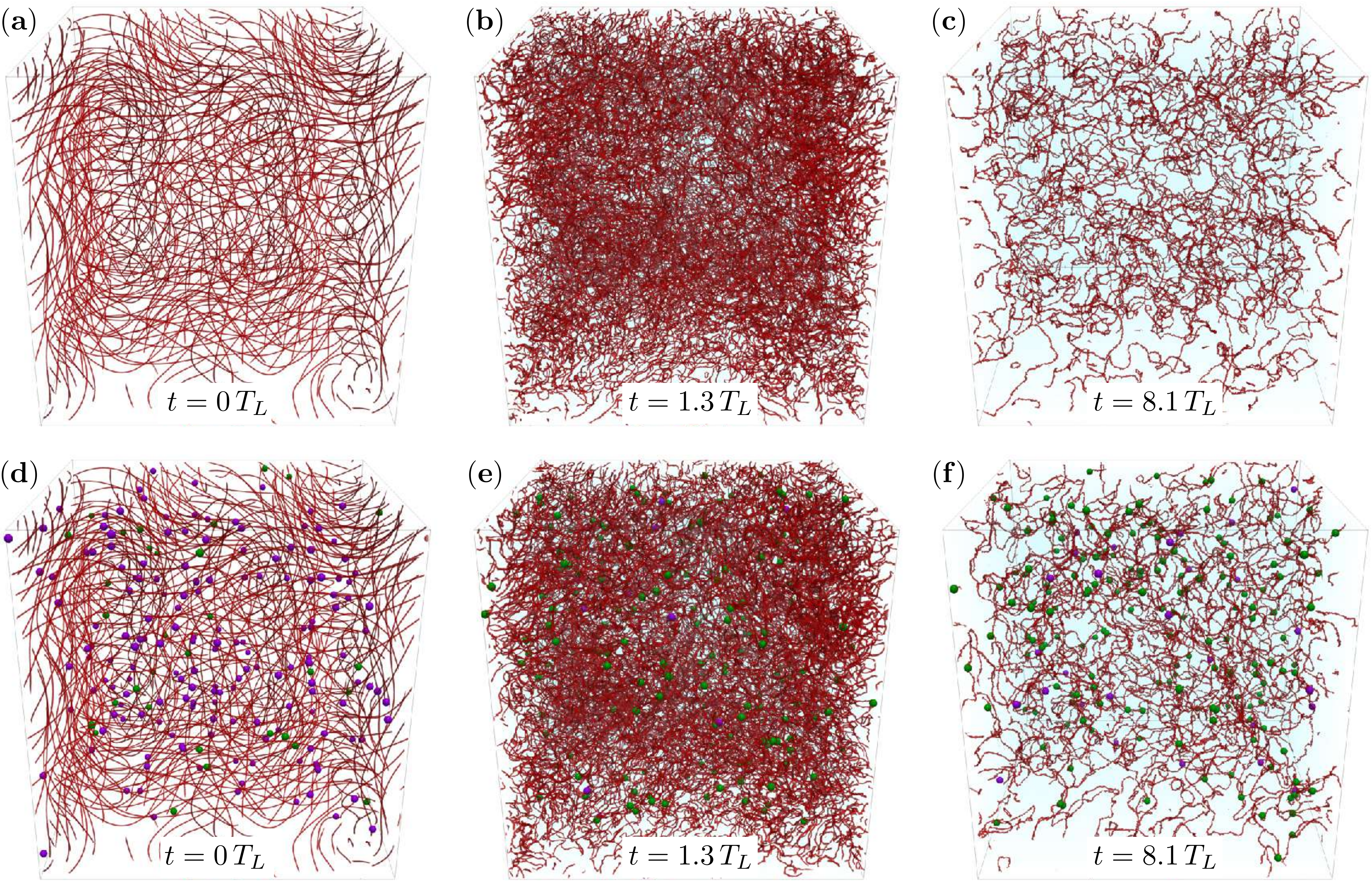}
\caption{(color online) Visualizations of the superfluid vortex tangle. Vortices are represented as isosurfaces in red of the density field ($\rho=0.15\rho_\infty$), sound is rendered in blue, trapped particles in green and free particles in purple. The upper row is without particles, the lower row is with $200$ neutrally buoyant particles of radius $\Rp = 4\xi$.   \textbf{(a}, \textbf{d)} The ABC initial states.  \textbf{(b}, \textbf{e)} The most turbulent regime ($t=1.3T_L$).  \textbf{(c}, \textbf{f)} A late time ($t=8.1T_L$). $T_L$ denotes the large-eddy-turnover time (see text).}
\label{Fig:3DVizFull}
\end{figure}

The first purpose of this work is to check whether and to what extent the presence of particles in the system modifies Kolmogorov turbulence. We add to the ABC initial condition a number of randomly distributed particles and let the system evolve under the dynamics (\ref{Eq:GPPfield}-\ref{Eq:GPPparticles}). In Fig.\ref{Fig:3DVizFull}.a, b and c the three stages of the evolution (respectively initial condition, turbulent vortex tangle and residual filaments in a bath of sound) are visualized in the case of $200$ neutrally buoyant particles of radius $4\xi$. 
Movies of this simulation and other with particles of a different size can be found in the Supplemental Material. Trapped particles by vortices are displayed in green, whereas free ones in purple. The algorithm to distinguish a trapped particle from a free one is based on the circulation around it and it is discussed in Section \ref{Subsec:MotionOfParticles}.

In Fig. \ref{Fig:3DVizFull} we observe that the building up and decay of the turbulent tangle is not strongly modified by the presence of particles. Moreover, it can be noticed how during the first stages of the evolution of the system the majority of particles gets trapped into the vortices. At zero temperature, as there is no normal component in the flow, no drag is experienced by the particles and their motion is completely driven by the pressure gradients. As a consequence, they are attracted by quantum vortices \cite{GiuriatoApproach,CaptureBerloff,ReviewNewcastle}. During the turbulent regime, violent and strongly nonlinear events like reconnections dominate the vortex dynamics and the flow evolution. A fundamental question is whether and how much the hydrodynamical attraction between vortices and particles is sufficient to keep them attached to the filaments. Indeed, since quantum vortices are actually the main actors of turbulence in superfluid, if particles are really able to follow them in this regime, it is a good indication that they are suitable to be used as probes.

In the next section, we will quantitatively study the effect of particles on quantum turbulent flows. We will first focus on the large scales of the flow, where Kolmogorov turbulence takes place. Then the particle dynamics and their statistics will be addressed.

\subsection{The effect of particles on Kolmogorov superfluid turbulence \label{SubSec:KolmogorovAndParticles}}

\begin{figure}[h!]
\includegraphics[width=.99\linewidth]{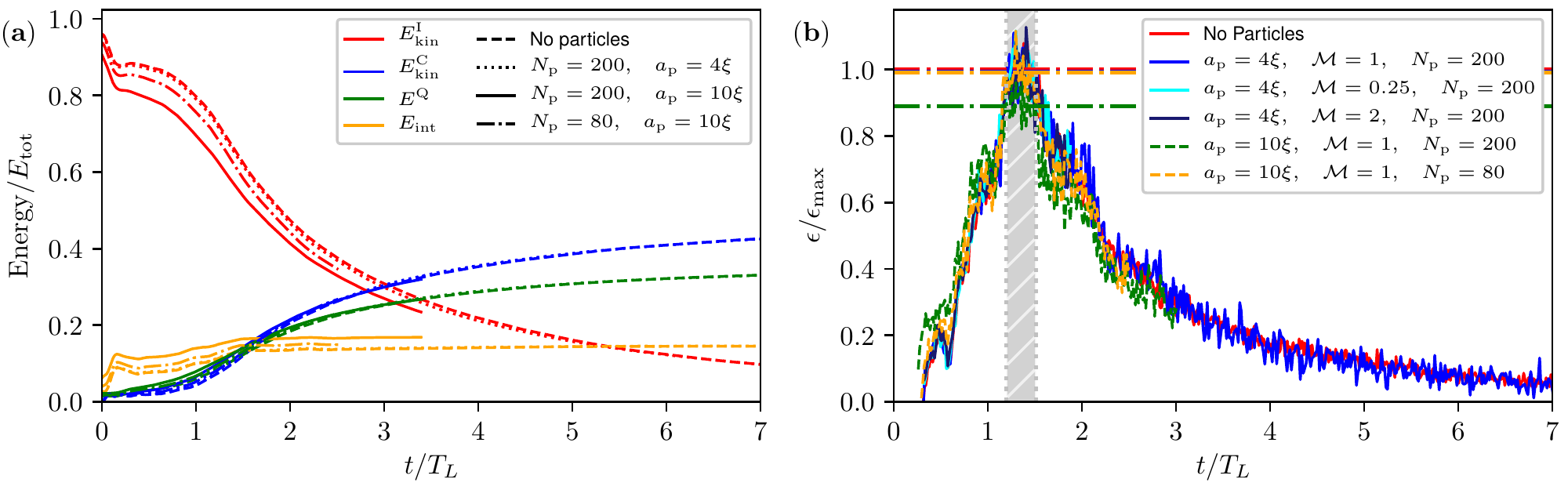}
\caption{ (color online) \textbf{(a)} Time evolution of the superfluid energy components in the cases with no particles (dashed line), $200$ small particles (dotted line),  $200$ large particles (solid line) and  $80$ large particles (dash-dotted line). \textbf{(b)} Incompressible energy dissipation rate for different number of particles with different sizes and different masses (solid lines). Dash-dotted horizontal lines of the corresponding colors indicate the value of the maximum of dissipation, obtained averaging over the shaded region. The dissipation is expressed in units of its maximum $\epsilon_\mathrm{max}$ in the case without particles.}
\label{Fig:enerdiss}
\end{figure}

We shall start our analysis by comparing the temporal evolution of global quantities. In Fig.\ref{Fig:enerdiss}.a the time evolution of the different components of the energy is displayed.
Times are expressed in units of the large-eddy-turnover time defined as $T_L=L/2v_\mathrm{rms}$, where $v_\mathrm{rms} = \sqrt{2E_\mathrm{kin}^\mathrm{I}(t=0)/3}$ is the root-mean-square velocity associated to the initial vortex tangle and $L/2$ is its characteristic length scale. We compare the case where no particles are present in the flow to the cases having particles of different sizes and of relative mass $\mathcal{M}=1$. 
The net transfer of incompressible energy towards compressible, quantum and internal energy is qualitatively unchanged in the various cases. The only difference is a slightly lower value of the incompressible energy in the case of large particles, in favor of the internal energy of the superfluid. Such effect is more evident if the number of large particles is increased, and could be related to an increment of the filling fraction $\Phi$, namely the fraction of the total volume occupied by the particles. In fact, for $N_\mathrm{p}=200$ particles of radius $\Rp=4\xi$ the filling fraction is $\Phi=0.1\%$, for $N_\mathrm{p}=80$ particles of radius $\Rp=10\xi$ it is $\Phi=0.8\%$ and for $N_\mathrm{p}=200$ particles of radius $\Rp=10\xi$ we have $\Phi=2.1\%$. The kinetic and repulsion energies of the particles, as well as the particle-vortex interaction $E_\mathrm{P}^\mathrm{GP}$ are negligible compared to the other energies throughout the duration of the simulations (data not shown).

The dissipation rate of the incompressible kinetic energy is reported in Fig.\ref{Fig:enerdiss}.b for particles of different masses and different sizes. The dissipation increases in the early stages, when the energy begins to be transferred to the smaller scales, it reaches a maximum when all the scales are excited, and then starts to decay since no forcing is sustaining the turbulence. We observe that the evolution of the dissipation is clearly not significantly modified by the presence of particles. In particular, the value of the maximum of dissipation, which is the signature of the most turbulent state reached by the tangle, is slightly lower only in the case where many large particles are moving in the system. In particular for this case, it is about $90\%$ of $\epsilon_\mathrm{max}$, the value measured in the case with no particles. The shaded region in Fig.\ref{Fig:enerdiss}.b represents the most turbulent time of the simulations. We consider that in this short stage, the system is in a quasi-steady state and we perform the temporal average of certain physical quantities in order to improve statistical convergence.

\begin{figure}[h!]
\includegraphics[width=.99\linewidth]{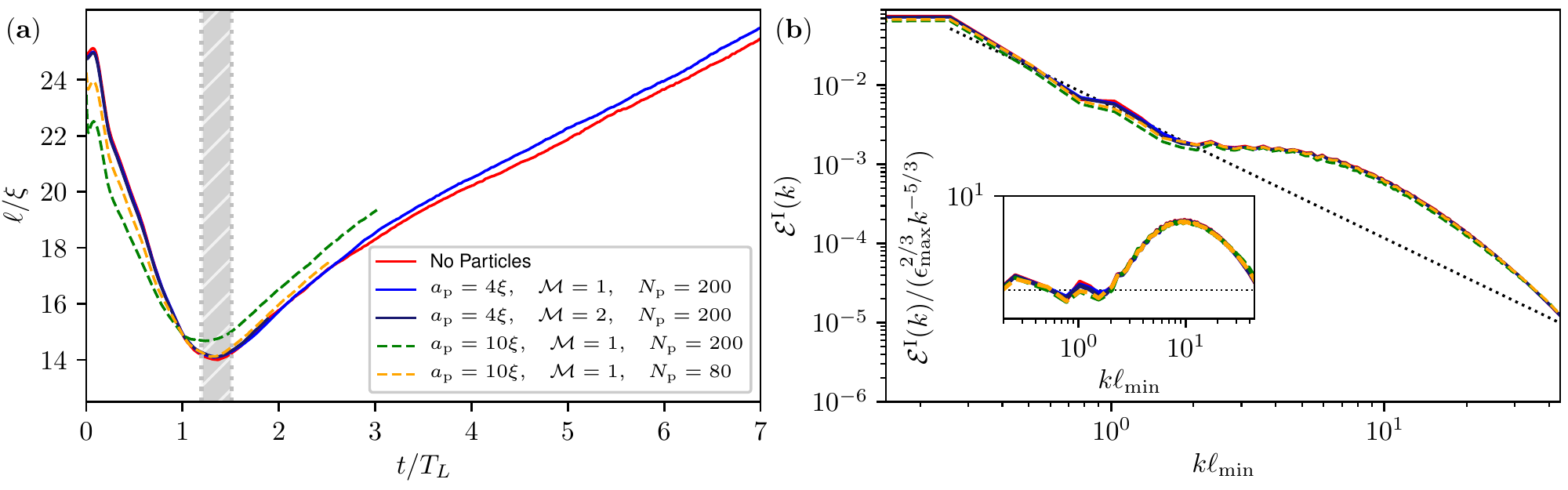}
\caption{(color online)  \textbf{(a)} Time evolution of the mean inter--vortex distance for different numbers of particles of different sizes and different masses. \textbf{(b)} Incompressible energy spectrum for different numbers of particles of different sizes and different masses. \textbf{(inset)} Compensated incompressible energy spectrum.  Solid lines refer to particles of size $\Rp=4\xi$, dashed lines refer to particles of size $\Rp=10\xi$. Dotted line is the classical scaling $\epsilon_\mathrm{max}k^{-5/3}$. The spectrum is computed averaging over times in the shaded region.}
\label{Fig:vortspeck}
\end{figure}

Two other important quantity that is not affected much by the interplay between tangle and particles is the mean inter--vortex distance $\ell$, whose time evolution is reported in Fig.\ref{Fig:vortspeck}.a.
The mean inter--vortex distance is then estimated as $\ell = \sqrt{V/L_\mathrm{v}}$, where $L_\mathrm{v}$ is the total vortex length in the system.
This latter is estimated using the method introduced in \cite{Noretal}, where $L_\mathrm{v}$ is shown to be related to the proportionality constant between the incompressible momentum density $J^\mathrm{I}(k)$ of the flow and the spectrum of a two-dimensional point-vortex $J_\mathrm{vort}^\mathrm{2D}(k)$:
\begin{equation}
\frac{L_\mathrm{v}}{2\pi} = \frac{\sum_k J^\mathrm{I}(k)}{\int J_\mathrm{vort}^\mathrm{2D}(k) \,\mathrm{d}k}.
\label{Eq:vlength}
\end{equation}
The spectra of momentum densities are the angle average of the norm in Fourier space of the momentum density $\mathbf{J} = \rho\mathbf{v}_\mathrm{s}$, and the incompressible part is obtained projecting onto the space of divergence-free fields. We have checked the validity of this formula by using the vortex filament tracking method described in \cite{VilloisTracking} at some checkpoints.

In the turbulent regime, where the dissipation gets its maximum, the total length of the entangled vortices is also larger by a factor $4$ compared to the initial condition, and the distance between the filaments is minimum. The value $\ell_\mathrm{min}\sim 14\xi$ of the inter--vortex distance in this regime will be used as a characteristic small length-scale of the Kolmogorov turbulent regime. Such length is smaller than the diameter of the largest particles considered ($2a_\mathrm{p}=20\xi$), but nevertheless this has no appreciable repercussions on the behavior of the observables studied. Furthermore, as shown in Fig.\ref{Fig:vortspeck}.c, the scaling of the incompressible energy spectrum $\mathcal{E}^\mathrm{I}(k)$ averaged around the maximum of dissipation is unaltered by particles in the system. The figure Fig.\ref{Fig:vortspeck}.b displays the incompressible energy spectrum. It is apparent that the scaling of the spectrum is always compatible with classical turbulence at scales larger than the inter--vortex distance, and the way in which the energy is accumulated at smaller scales is not modified by the particles. In the inset of Fig.\ref{Fig:vortspeck}.b the spectrum is compensated by with the Kolmogorov prediction $\mathcal{E}^\mathrm{I}(k) = C\epsilon_\mathrm{max}^{2/3} k^{-5/3}$ for classical hydrodynamic turbulence. The dotted horizontal black line shows that the value of the constant $C$ in the Kolmogorov law is a number of order $1$ for superfluid turbulence. 

\begin{figure}[h!]
\includegraphics[width=.99\linewidth]{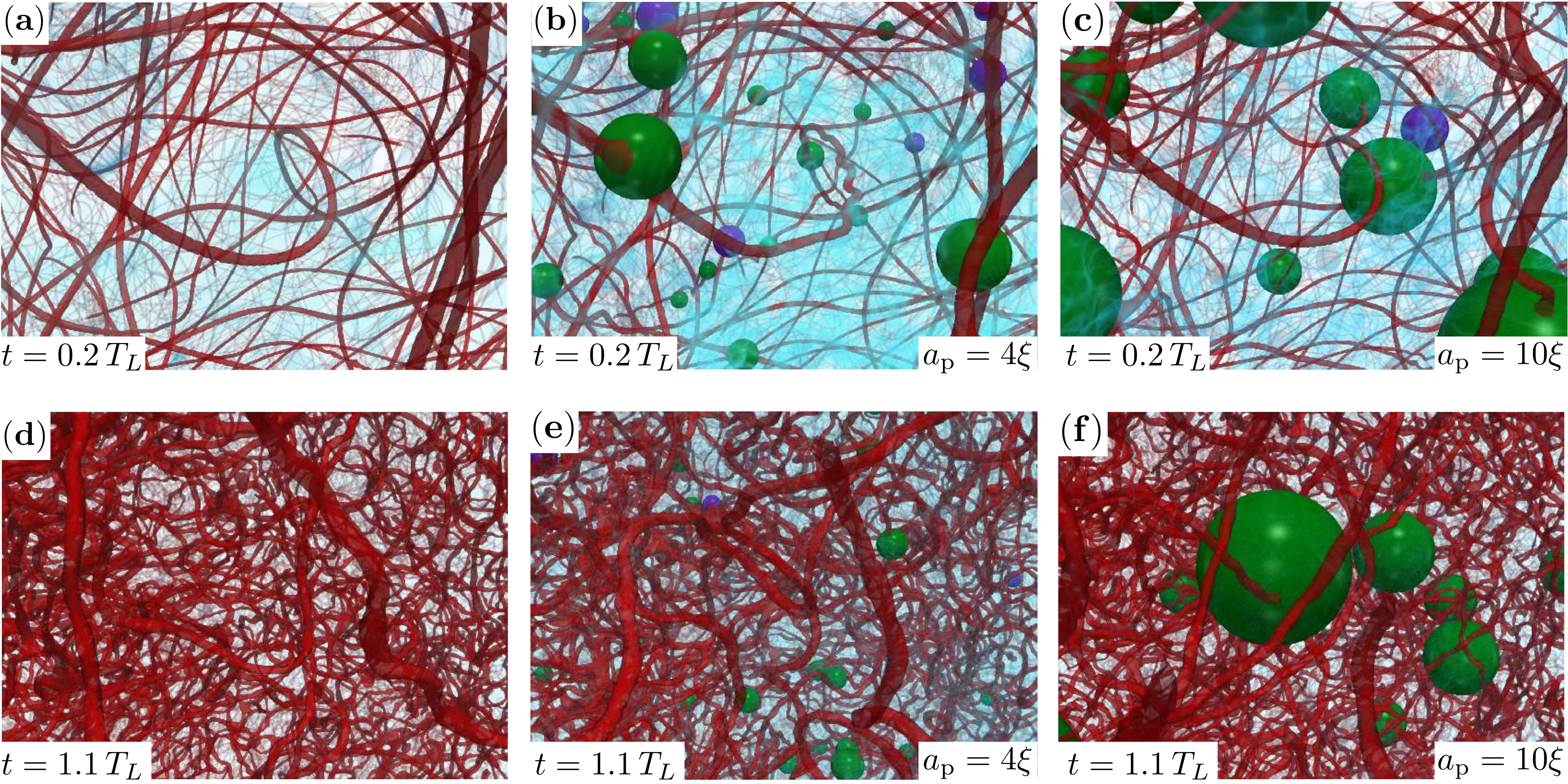}
\caption{(color online) Close-up of the superfluid vortex tangle at the early stage of the simulation (upper row: $t=0.27T_L$) and during the turbulent regime (lower row: $t=1.1T_L$) for the cases with no particles (left column \textbf{(a)}, \textbf{(d)}), small particles (central column \textbf{(b)}, \textbf{(e)}: $\Rp=4\xi$) and large particles (right column \textbf{(c)}, \textbf{(f)}: $\Rp=10\xi$). Vortices are represented as isosurfaces of the density field ($\rho=0.15\rho_\infty$) and rendered in red, sound is rendered in blue, trapped particles in green and free particles in purple.}
\label{Fig:zoom}
\end{figure}

The only appreciable difference observed between the case with and without particles is that in the early stages of the evolution, the trapping of particle perturbs the vortex filaments and excite Kelvin waves. A comparison between the volume renderings can be seen in the upper row of Fig.\ref{Fig:zoom}.
Such perturbations propagate during the evolution of the tangle. At the times when turbulence is developed, the details of the vortex configurations are completely different (see lower row of Fig.\ref{Fig:zoom}). Nevertheless, the statistical properties of the system in this regime remain unchanged. We stress that the inter-vortex distance in quantum turbulence experiments lies typically in the range $10-100\,\mu m$, which is equal or slightly larger than the particle size \cite{PaolettiVelStat2008,LaMantiaVelocity,LaMantiaAcceleration}. In this sense, the simulations presented here are compatible with the experimental parameters. They thus support the believe that active particles have effectively no influence on the typical development and decay of quantum turbulence. This numerical fact helps to validate past and future experiments that use particles as probes of superfluids. 

On the other hand, because of the lack of a Stokes drag in the system, particles cannot be treated as simple tracers of the superfluid velocity $\mathbf{v}_\mathrm{s}$. Nevertheless, if they remain trapped inside the vortices they can track the evolution of the vortex filaments, which are the structures that effectively become turbulent. With the purpose of characterizing this scenario, in the next section we investigate the motion of particles once they are immersed in a tangle of quantum vortices.

\subsection{Motion of particles in the superfluid vortex tangle\label{Subsec:MotionOfParticles}}

Looking at the time evolution of the vortex tangle (see Fig.\ref{Fig:3DVizFull} and movies in the Supplemental Material), the first thing that is apparent is how particles quickly get trapped into vortex filaments. This dynamics is expected and it has been studied in the case where vortices move slowly \cite{GiuriatoApproach}. It is a consequence of the pressure gradients. However, it is less obvious if such behavior remains dominant when turbulence take place and reconnections become frequent. 

\begin{figure}[h!]
\includegraphics[width=.99\linewidth]{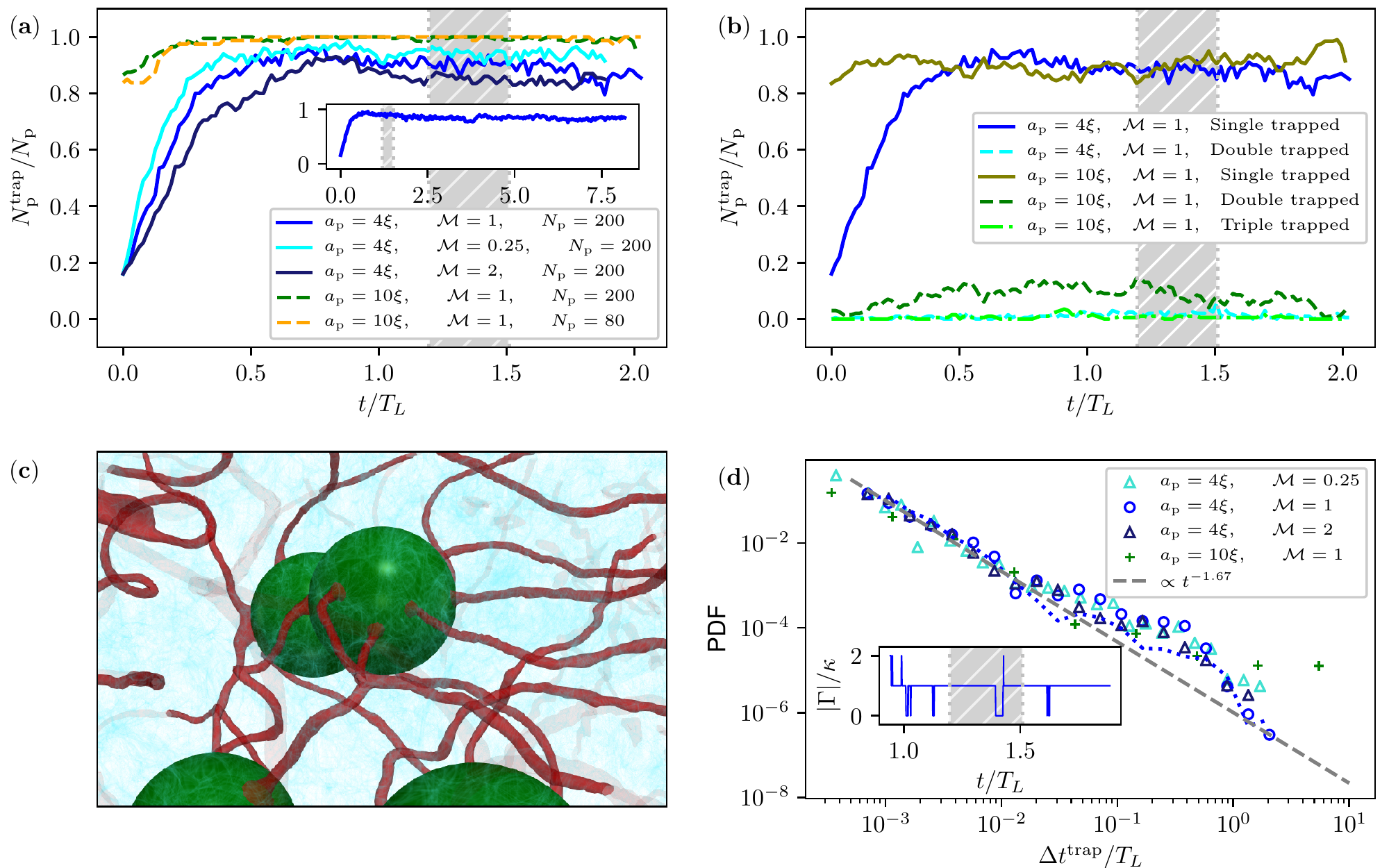}
\caption{(color online) \textbf{(a)} Fraction of trapped particles as a function of time for different numbers of particles of different sizes and different masses. \textbf{(inset)} The same for longer time in the case of $200$ neutrally buoyant particles of size $\Rp=4\xi$. \textbf{(b)} Comparison between the fraction of multiply trapped particles as a function of time for neutrally buoyant particles. \textbf{(c)} Volume rendering of large particles ($a_\mathrm{p}=10\xi$) multiply trapped by quantum vortices. Vortices are rendered in red, sound in blue, particles in green. \textbf{(d)} Probability density function of the continuous time spent by particles inside vortices for different species of particles. Dotted blue line corresponds to the same simulation of blue circles (particles with size $a_\mathrm{p}$ and mass $\mathcal{M}=1$) but averaged over the full simulation times.)\textbf{(inset)} Absolute value of the circulation around a single particle of size $a_\mathrm{p}=4\xi$ and mass $\mathcal{M}=1$ as function of time. The PDF is computed averaging over times in the shaded region.}
\label{Fig:trap}
\end{figure}

We study the evolution of particles and compute wether they are free or trapped by vortices. The temporal evolution of the fraction of trapped particles is displayed in Fig.\ref{Fig:trap}.a for all runs.
This measurement is made by computing the circulation $\Gamma = \oint_\mathcal{C} \mathbf{v}_\mathrm{s}\cdot\mathrm{d}\mathbf{x}$ of the superfluid velocity $\mathbf{v}_\mathrm{s}$ along contours $\mathcal{C}$ encircling each particle, and counting for which particles it is different from zero. Specifically, we compute the circulation along many parallel square contours of side $2(\Rp +\Delta_x)$ around each particle, where $\Delta_x$ is the grid spacing. If the circulation around at least one of these contours is different from zero, the particle is considered as trapped \footnote{The circulation measured with this method is subjected to a numerical error coming from the grid spacing. Such error is removed in post-processing, knowing that $\Gamma$ can only be an integer multiple of $\kappa$. Furthermore, extremely high values of $\Gamma$ have been removed since they are related to the ill-defined situation in which a topological defect is placed at the boundary $\mathcal{C}$.}. For practical reasons, due to the parallelization of the numerical code, we consider only contours perpendicular to the $z$ axis of the computational box. As a consequence, the protocol is not able to grasp vortices that are crossing the particles exactly on a plane perpendicular to the $z$ axis. This means that our estimation of the fraction of trapped particles is effectively a lower bound. However, it should be noticed that this pathological situation is an extremely rare situation that does not change the conclusions of our analysis. 

In the initial condition the particles are placed randomly in the computational box, it happens then that some of them are already positioned inside a vortex. In the case of particles with a size comparable to the inter--vortex distance the majority of particles is in this situation. In the first stages of the evolution of the flow, the number of trapped particles increases rapidly, until it gets stationary always at times much smaller than one $T_L$.  The time needed to reach a stationary state slightly depends on the mass of the particles, as well as the fraction of trapped particles once a steady regime is reached. The steady value of $N_\mathrm{p}^\mathrm{trap}/N_\mathrm{p}$ is between $80\%$ and $90\%$ for small particles ($2a_\mathrm{p}<\ell$), while in average the totality of particles of size $2a_\mathrm{p}\sim \ell$ is found to be trapped by vortices, independently of the filling fraction. When the system reaches the most turbulent regime (indicated by the shaded region), the fraction of trapped particles does not undergo any appreciable changing. In the inset of Fig.\ref{Fig:trap}.a, $N_\mathrm{p}^\mathrm{trap}/N_\mathrm{p}$ is also shown for late times in the case of small particles of relative mass $\mathcal{M}=1$. It manifestly remains stable. This means that even when the density of vortex lines is decaying (along with the intensity of turbulence) the particles stay trapped inside vortices. Note that in this work we are dealing with homogenous and isotropic decaying quantum turbulence at low temperature. We mention that the fraction of trapped particles measured in thermal counterflow simulated by means of the VF method is lower that the one observed here \cite{BarenghiTracers}.

The circulation around each superfluid vortex filament is equal to a single quantum of circulation $\kappa$. As a consequence, measuring the circulation along a closed line $\mathcal{C}$ allows us to count the number of filaments in the region delimited by the line, provided that the quanta of circulation around every filament have the same sign. This is true also if the vortices are trapping particles, because their topological nature does not change. In Fig.\ref{Fig:trap}.b we show again the fraction of trapped particles, but now separating the number of particles trapped by multiple vortices. It turns out that at least the $5-10\%$ of the particles with size $2a_\mathrm{p}\sim\ell$ are always attached to at least two different filaments. Sometimes even more vortices pass simultaneously through the same particle, as it can be visualized in the volume plot of Fig.\ref{Fig:trap}.c. 

Once a particle is trapped by a vortex, it can experience violent events, for instance during vortex reconnections. In such circumstances, such a particle could be detached and expelled from the vortex until it will eventually get trapped by another vortex of the tangle. We compute the probability density function (PDF) of the continuous time intervals $\Delta t_\mathrm{trap}$ spent by the particles inside vortices regime. The PDFs for particles of different sizes and masses are displayed in Fig.\ref{Fig:trap}.d.
For all the species of particles examined, the probability distribution seems to follow roughly a power-law scaling in time $\sim {(\Delta t_\mathrm{trap})}^{-\alpha}$, with $\alpha\sim 1.67$. The PDF certainly vanishes much slower than an exponential decay at large $\Delta t^\mathrm{trap}$, which would typically result form a standard escape problem over energy barriers.  We checked that the intermittency of the circulation and the shape of the trapping time PDF are not peculiar of the most turbulent regime, since they persist also at the late times of the simulations (see dotted blue line in Fig.\ref{Fig:trap}.d).
Therefore, many particles spend a time at least of the order of the simulation time ($\sim 10 T_L$) inside a vortex filament, i.e. the typical escape time from the vortices is virtually infinite. This observation is exemplified in the inset of Fig.\ref{Fig:trap}.d, where the evolution of the circulation around a single small neutral particle is reported (the qualitative behavior is the same for the other particles). It is also clear that the time spent by the particles with zero circulation around them (namely free from vortices) are short. 
Since we established that particles immersed in a tangle spend most of the time inside vortex filaments, in the following we study their motion once they get trapped. 

\begin{figure}[h!]
\includegraphics[width=.99\linewidth]{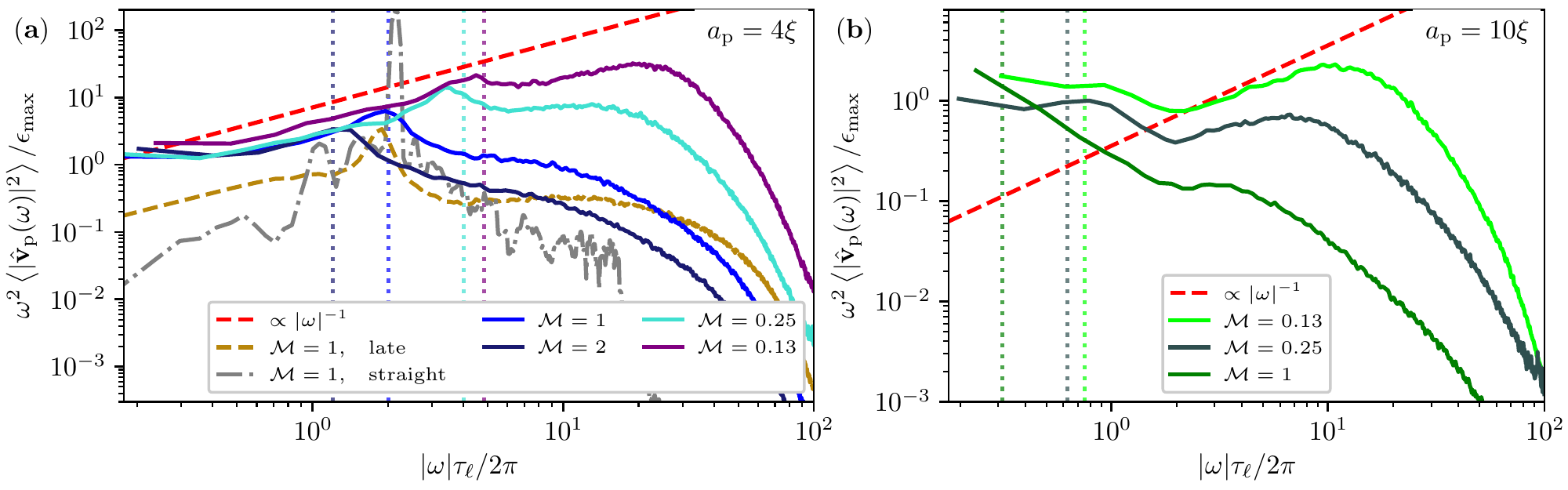}
\caption{(color online) Frequency spectrum of the particle velocity for particles of different masses and different sizes, compensated with the prediction for the Lagrangian spectrum in classical turbulence $\propto\epsilon/\omega^{2}$: \textbf{(a)} small particles with $\Rp=4\xi$; \textbf{(b)} large particles with $\Rp=10\xi$. Dash-dotted gray line is the frequency spectrum of a single small particle trapped in a straight vortex slightly perturbed. Dotted lines of corresponding colors are the prediction for the particle natural frequency $\Omega_\mathrm{p}$. Dashed red line is the scaling due to vortex reconnection  or Kelvin waves $\propto|\omega|^{-1}$. Dashed golden line is the spectrum evaluated at late times in the simulation ($6T_L<\tau<7T_L$).}
\label{Fig:speck}
\end{figure}

At large scales, the vortex tangle seems to behave as a classical hydrodynamic turbulent system. Therefore the first natural question is whether the particles can trace such large-scale fluctuations. In classical turbulence, it is well known that the Lagrangian velocity spectrum scales as 
\begin{equation}
\left\langle|\hat{\mathbf{v}}_\mathrm{p}(\omega)|^2\right\rangle=B\epsilon\omega^{-2},\label{Eq:LangVelSpec}
\end{equation}
 where  $B$ is a constant of order unity and $\hat{\mathbf{v}}_\mathrm{p}(\omega)$ is the Fourier transform of the Lagrangian particle velocity $\mathbf{v}_\mathrm{p}(t)$ \cite{YeoungLagrangian,tennekes}. Such scaling is valid in the inertial range $2\pi/T_L\ll\omega\ll2\pi/\tau_\eta$, where $\tau_\eta$ is the Kolmogorov time scale. In our case, we build an analogous of the Kolmogorov time scale under the assumptions that the dissipation rate $\epsilon_\mathrm{max}$ is the only important physical parameter in the classical turbulence regime and that the Kolmogorov turbulent cascade ends at the inter--vortex distance $\ell_\mathrm{min}$ . Therefore, we define the smallest time scale of the classical turbulence regime as $\tau_\ell=\left(\ell_\mathrm{min}^2/\epsilon_\mathrm{max}\right)^{1/3}$, and we expect classical turbulent phenomenology to hold for times $\tau_\ell\ll t\ll T_L$. 
In Fig.\ref{Fig:speck} the measurement of the frequency spectrum of the particle velocity $\left\langle|\hat{\mathbf{v}}_\mathrm{p}(\omega)|^2\right\rangle = \left\langle|\int \dot{\mathbf{q}}(t) e^{-i\omega t}\,\mathrm{d}t|^2\right\rangle$ during the turbulent regime is shown for different species of particles, compensated with the classical scaling $\epsilon_\mathrm{max}\omega^{-2}$. 
Note that the average which defines the spectrum is meant over different realizations. In numerics we average over all the particles trajectories during the turbulent regime. At frequencies $\omega<\tau_\ell/2\pi$ the spectra approach a plateau of value one, confirming that particles sample well the flow and their behavior is described by the standard classical turbulence picture at large scales. Note that the classical temporal inertial range of our simulations is pretty small, since $T_L\sim 5 \tau_\ell$.
For comparison, we also present the velocity spectrum of a particle of size $\Rp=4\xi$ and mass $\mathcal{M}=1$, computed in a temporal window at much later times, when Kolmogorov turbulence has decayed and only few vortices are left. Note that a $\omega^{-2}$ scaling of the Lagrangian velocity spectrum has been also observed in numerical simulations of the vortex filament model   \cite{BarenghiTangle}, although not in the Kolmogorov inertial range and not related to the energy dissipation rate nor to Kolmogorov turbulence.

As expected, in our simulations no Kolmogorov scaling is observed at small time scales. Indeed, one of the most striking features of quantum turbulence is the crossover between the classical Kolmogorov regime and the physics taking place at scales smaller than the mean inter--vortex distance. Unlike classical turbulence (see for instance \cite{YeoungLagrangian}), there is still a non trivial scaling at time scales shorter than $\tau_\ell$. Such difference is a consequence of the quantum nature of the system, here manifested by the presence of quantized vortices.

When a particle is trapped by a vortex, the superfluid flow turns around it. As a consequence, while the particle moves, it experience a Magnus force. This lift force is simply expressed as $\mathbf{F}_\mathrm{Magnus}=\frac{3}{2}\rho_\infty a_\mathrm{p} \mathbf{\Gamma}\times(\dot{\mathbf{q}}-\mathbf{v}_\mathrm{s})$, where the circulation vector $\mathbf{\Gamma}$ is oriented along the vortex filament and the superfluid velocity $\mathbf{v}_\mathrm{s}$ contains the contributions of the mean flow and the vortex motion \cite{GiuriatoCrystal, magnus_sphere}. Magnus effect induces a precession of the particle about the filament with the characteristic angular velocity 
\begin{equation}
\Omega_\mathrm{p}=\frac{3}{2}\frac{\rho_\infty a_\mathrm{p}}{M_\mathrm{p}^\mathrm{eff}}\Gamma,
\label{Eq:omegaP}
\end{equation}
where the particle effective mass $M_\mathrm{p}^\mathrm{eff} = M_\mathrm{p}+\frac{1}{2}\mathrm{M}_0 = (\mathcal{M}+\frac{1}{2})M_0$  takes into account the added mass effect due to the mass of the superfluid displaced by the particle $M_0$. As mentioned in \cite{GiuriatoCrystal}, for current experiments with hydrogen particles in superfluid helium, this frequency is of order $10-100$Hz.
If the Magnus force is the main force acting on a trapped particle, the Newton equation $M_\mathrm{p}^\mathrm{eff}\ddot{\mathbf{q}}=\mathbf{F}_\mathrm{Magnus}$ implies the following expression for the frequency spectrum of the particle velocity:
\begin{equation}
\left\langle|\hat{\mathbf{v}}_\mathrm{p}(\omega)|^2\right\rangle = \frac{\Omega_\mathrm{p}^2}{\Gamma^2(\omega-\Omega_\mathrm{p})^2} \left\langle|\mathbf{\Gamma}\times\hat{\mathbf{v}}_\mathrm{s}(\omega)|^2 \right\rangle.
\label{Eq:velspeck}
\end{equation}
Independently of the external superfluid velocity, the expression (\ref{Eq:velspeck}) predicts that the spectrum $\left\langle|\hat{\mathbf{v}}_\mathrm{p}(\omega)|^2\right\rangle$ must be peaked around the natural frequency of trapped particles $\omega=\Omega_\mathrm{p}$. Such behavior has been studied in detail in the case of particles trapped inside slightly perturbed straight vortex filaments {\cite{GiuriatoCrystal}. The spectrum of this simple configuration is also reported for comparison in Fig.\ref{Fig:speck}.a for a small particle of relative unit mass. A clear bump in the frequency spectrum, corresponding to $\Omega_\mathrm{p}$, is still visible when particles are immersed in a complex quantum vortex tangle. For the large particles, the presence of a peak is less evident because the natural frequency is lower, and therefore a longer sampling (in time) would be necessary to resolve it properly ($2\pi/\Omega_\mathrm{p} = 0.7T_L$ for the particles of size $a_\mathrm{p}=10\xi$ and mass $\mathcal{M}=1$). Moreover, as large particles are multiply trapped by many vortices, the resulting motion is certainly more complex than a precession with a single characteristic angular frequency of one single vortex. The broadness of the peak around the Magnus frequency for the small particles in Fig.\ref{Eq:omegaP}.a, could be also related to this fact.

At small time scales, a different scaling of the velocity spectrum is observed for the light particles, now in agreement with $\left\langle|\hat{\mathbf{v}}_\mathrm{p}(\omega)|^2\right\rangle \propto |\omega|^{-1}$. This behaviour is consistent with the fact that at scales smaller than the inter--vortex distance, the typical velocities of a superfluid turbulent tangle are supposed to scale as $v_\mathrm{\rm fast}(t)\propto\sqrt{\kappa/|t-t_0|}$, because the circulation becomes the only relevant physical parameter and the motion of vortices is dominated by their mutual advection and reconnections. In this scenario, if particles are sufficiently light to be able to follow the fast vortex dynamics, we can substitute $\left\langle|\hat{\mathbf{v}}_\mathrm{p}(\omega)|^2\right\rangle \sim \hat{v}_\mathrm{\rm fast}^2(\omega) \propto \kappa|\omega|^{-1}$. Another effect that could contributes to the same result is the attraction of particles by the vortices, since the scaling in time of the particle-vortex distance is the same of vortex reconnection \cite{GiuriatoApproach}. Note that for the heaviest particles such fast scaling is absent, since their reaction is probably too slow to be sensible to the fast fluctuations of the tangle.

\subsection{Particle velocity and acceleration statistics \label{Subsec:PartStatistics}}

Unlike classical turbulence, where the statistics of the one-point particle velocity $v$ is known to be Gaussian \cite{frisch1995turbulence}, experiments in superfluid helium using hydrogen and deuterium particles as tracers have reported long tails, with a $v^{-3}$ power--law scaling in their velocity distribution \cite{PaolettiVelStat2008,LaMantiaVelocity,LaMantiaQT}. Such scaling has been related to the singular velocity field of quantized vortices \cite{barenghiPDF,paoletti_review}. At low temperatures, as Stokes drag is negligible, particles should not move with the superfluid flow and such scaling can be understood as a consequence of quantum vortex reconnections sampled by trapped particles \cite{PaolettiVelStat2008,ReconnectionGiorgio}. Furthermore, in reference \cite{LaMantiaVelocity}, by using particle tracking velocimetry in counterflow turbulence, it was shown that while varying the sampling scale, the velocity PDFs continuously change from Gaussian statistics to power-law tails, the crossover taking place at scales of the order of the inter--vortex distance. In this last section we present measurements of particle velocity and acceleration statistics within the GP-P model. 

We start the discussion by presenting the Eulerian velocity field. Formally, the velocity of the superfluid is simply given by $\nabla \phi$. This field contains the density fluctuations, as well as the divergence of the vortex velocity flow close to its core. This divergence leads to the well observed $v^{-3}$ scaling of velocity PDF  \cite{barenghiPDF,BaggaleyBarenghiVelStat,ShuklaNJP}. The PDF of $\nabla\phi$ is displayed in Fig.\ref{Fig:PDFv}.
\begin{figure}[h!]
\includegraphics[width=.99\linewidth]{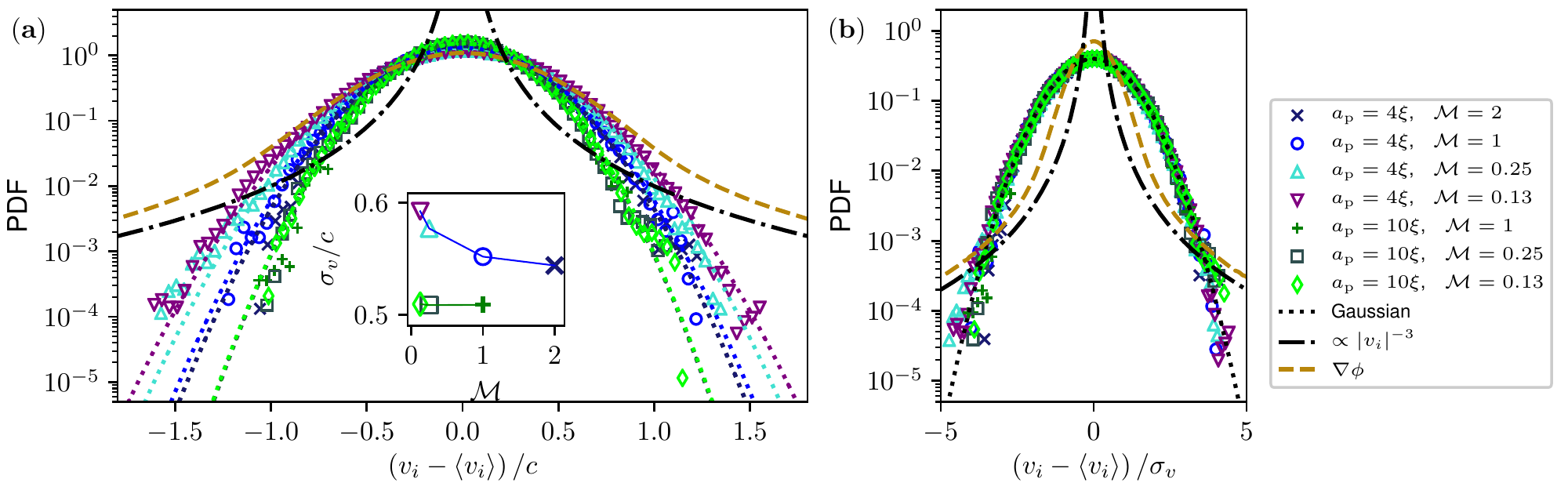}
\caption{\textbf{(a)} Probability density function of the single component particle velocity, for different species of particles. Dotted golden line the Eulerian velocity field $\nabla\phi$, corresponding to the simulation without particles at the time $1.4\,T_L$. The data for the particles are averaged in time between $t=1.2T_L$ and $t=1.6T_L$. \textbf{(inset)} Standard deviation of the particle velocity as a function of the particle mass. \textbf{(b)} The same of \textbf{(a}) but with the velocities normalized by the standard deviation $\sigma_v$.  Dotted lines are gaussian, dash-dotted line is a power-law scaling $|v_i|^{-3}$.}
\label{Fig:PDFv}
\end{figure}
We turn now to analyze the particle velocity PDFs. We compute the velocity PDFs for all runs, in the turbulent regime. Data is  filtered with a Gaussian convolution in order to smooth out the noisy oscillations at frequencies $\omega<\omega_\mathrm{noise}=50\,(2\pi/\tau_\ell)$ (see Appendix \ref{AppComp}). In Fig.\ref{Fig:PDFv} the PDF of the single component velocity is plotted for all the species of analyzed particles. In Fig.\ref{Fig:PDFv}.a, velocities are expressed in terms of the speed of sound $c$, whereas in Fig. \ref{Fig:PDFv}.b they are normalized by their root--mean--squared values. The root--mean--squared values are displayed in the inset of Fig.\ref{Fig:PDFv}.a as a function of the mass for the two particle sizes. It is apparent from Fig.\ref{Fig:PDFv}.b, that the particle statistics exhibits a Gaussian distribution. Note that Gaussian velocity statistics were also observed in thermal counterflow simulations of the vortex filament method with tracers particles \cite{BarenghiTracersFirst}. The absence of power-law tails could be a consequence of weak statistical sampling of large velocity fluctuations due to the low number of particles present in the system and/or by compressible effects of the GP model. 
We will comment more about this in the Discussion section.

We would like to remark here that high frequency fluctuations are strongly sensitive to numerical artifacts. In the Appendix, inspired by the experimental results of reference \cite{LaMantiaVelocity}, we have computed the velocity PDFs of the velocity fluctuations filtered at a given frequency $\omega_c$. The frequency was varied from values lower to larger than $2\pi/\tau_\ell$. For one simulation we have compared two different interpolation methods to evaluate the force term in Eq.\eqref{Eq:GPPparticles} needed to drive the particles. It turns out that for the fourth--order B--spline method, the velocity PDFs start to develop tails while the filtering scale is varied, that eventually lead to a $v^{-3}$ scaling. However, when using Fourier interpolation, that is an exact evaluation (up to spectral convergence of the pseudo--spectral code) of the force term, the PDFs do not develop any tail and remain Gaussian. We have decided to keep this example with spurious numerical effects in the Appendix, as it might be useful for future numerical studies and data analysis of similar problems. We have checked that the results presented in the paper are independent of the interpolation scheme.

We turn now to study the acceleration statistics. As displayed in Fig.\ref{Fig:PDFa}.a, the PDF of the acceleration presents some deviations from a Gaussian distribution at large values.
\begin{figure}[h!]
\includegraphics[width=.99\linewidth]{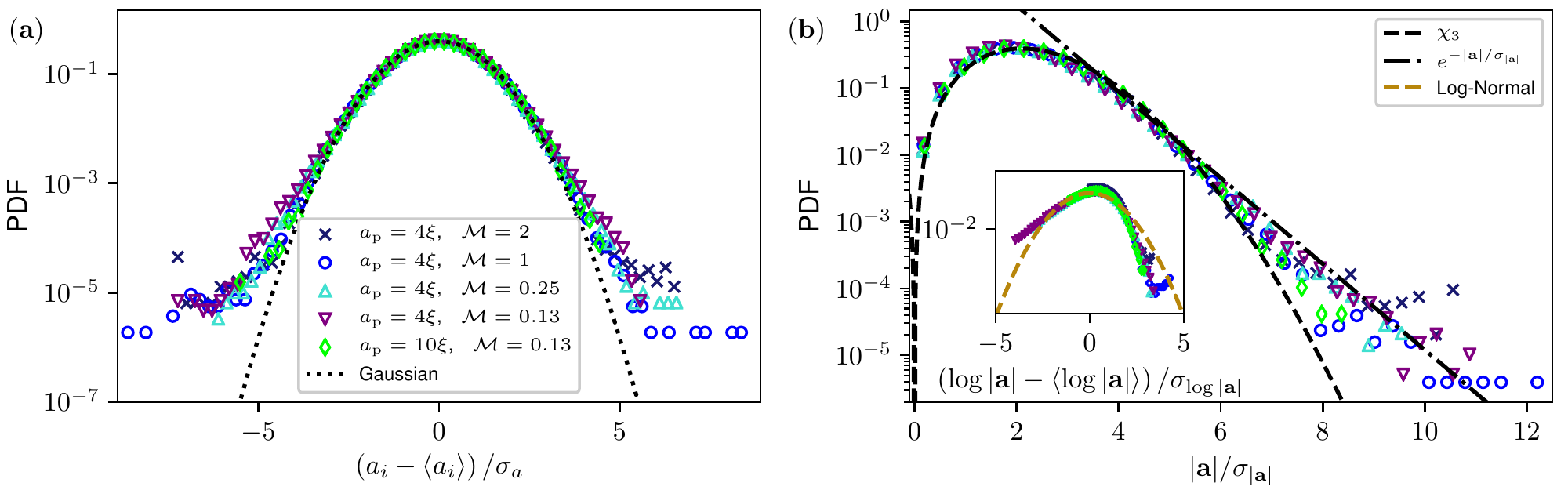}
\caption{(color online) \textbf{(a)} Probability density functions of the single component particle acceleration. \textbf{(b)} Probability density functions of the norm of the particle acceleration. Dotted line is a gaussian, dashed line is a $\chi_3$ distribution and dash-dotted line is an exponential tail $e^{-|\mathbf{a}|/\sigma_{|\mathbf{a}|}}$. \textbf{(inset)} Probability density functions of the logarithm of the norm of the particle acceleration. Dashed golden line is a log--normal distribution.}
\label{Fig:PDFa}
\end{figure}
The norm of the acceleration has also an exponential tail for $|\mathbf{a}|>\sigma_{|\mathbf{a}|}$, as displayed in Fig.\ref{Fig:PDFa}.b. The core of the PDF in this case is a $\chi_3$ distribution, which is expected for the norm of a vector with Gaussian components. In classical Lagrangian turbulence, the norm of the particle acceleration is observed to obey a log--normal distribution \cite{MordantAcceleration}. In the inset of Fig.\ref{Fig:PDFa}.b, we compare our data with such distribution. For the lightest and smallest particle, the small accelerations appears to be more probable than in the classical case. Note that, as pointed out in \cite{MordantAcceleration}, small values of the acceleration are very sensible to experimental (numerical) errors. By contrast, the large accelerations are less probable than a log-normal distribution. This observation is compatible with classical numerical calculations in the framework of the viscous vortex filament model, in which it has been shown that, because of inertia, solid particles undergo less rapid changes of velocity than fluid particles \cite{VFP}.

Finally, in Fig.\ref{Fig:CorrA}, we show the two--point correlator of the particle acceleration, defined as:
\begin{equation}
\rho^a(t) = \frac{ \left\langle a_i(t_0)a_i(t_0+t) \right\rangle - \left\langle a_i(t_0) \right\rangle \left\langle a_i(t_0+t) \right\rangle } {\sigma_a(t_0)\sigma_a(t_0+t)}.
\label{Eq:corr}
\end{equation}
\begin{figure}[h!]
\includegraphics[width=.99\linewidth]{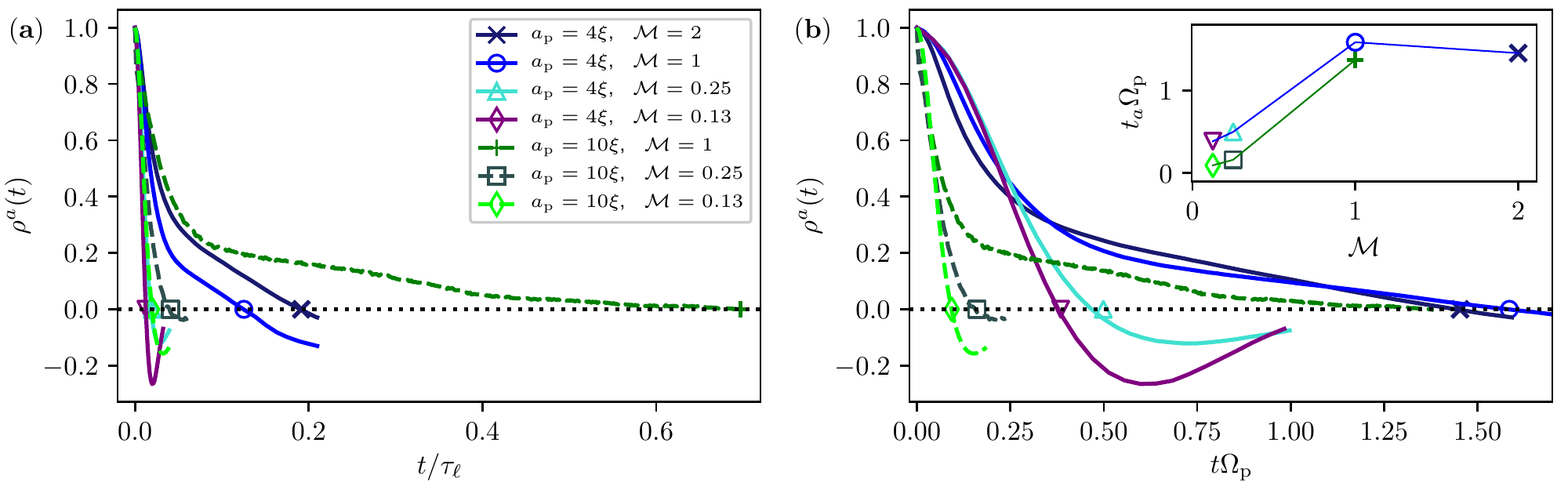}
\caption{(color online) Acceleration two--point correlator, plotted versus time normalized by the dissipation time scale $\tau_\ell$ \textbf{(a)}, and by the Magnus natural frequency. $1/\Omega_\mathrm{p}$ \textbf{(b)} Markers indicate the time of acceleration decorrelation $t_a$. \textbf{(inset)} $t_a$ normalized by $1/\Omega_\mathrm{p}$ as a function of the particle relative mass.}
\label{Fig:CorrA}
\end{figure}
In classical Lagrangian turbulence, the decorrelation time $t_a$ (such that $\rho^a(t_a)=0$) is related to the Kolmogorov time scale $t_a=2\tau_\eta$ \cite{YeungPopeLong}. This is not the case in quantum turbulence. Figure  \ref{Fig:CorrA}.a displays the autocorrelation $\rho^a(t)$ for all the simulations. It is apparent that the acceleration decorrelates much faster than $\tau_\ell$, the equivalent of the Kolmogorov time scale in our system. This fact is a consequence of the myriad of physical phenomena taking place at smaller scales. As most particles are trapped by vortices, they oscillate at the Magnus frequency $\Omega_\mathrm{p}$ in Eq.\eqref{Eq:omegaP}. If time is normalized by $\Omega_\mathrm{p}$ (\ref{Eq:omegaP}), then $t_a\Omega_\mathrm{p}$ becomes of order $1$, at least for the heaviest particles (see Fig.\ref{Fig:CorrA}.b and the inset therein). For the lightest particles the decorrelation time is even lower, meaning that they are sensible to other mechanisms, like reconnection events between vortex filaments and Kelvin waves excitations at even smaller scales.

\section{Discussion \label{Sec:Discussion}}

In this work we used the Gross-Pitaevskii model to study free decaying quantum turbulence at zero temperature in presence of finite size active particles. We considered different families of spherical particles having sizes smaller than and of the order of the mean inter-vortex distance. We first performed a standard analysis of the observables commonly used for studying Kolmogorov turbulence, such as the energy decomposition, the temporal evolution of mean energy, the rate of incompressible kinetic energy and the mean inter-vortex distance. Although particles are active and get captured by vortices generating Kelvin waves, there is not a significant impact at scales larger than the inter-vortex distance, where Kolmogorov turbulence takes place. Monitoring the motion of the particles in the system, we confirmed their tendency to remain trapped into vortex filaments during the evolution of the tangle, with intermittent episodes of detachment and recapture. This behavior is independent of the vortex line density. We also found that particle can be easily captured simultaneously by several quantum vortices.

We also studied turbulence from the Lagrangian point of view. In particular, we computed the power spectra of the particle velocities. At large scales the particle dynamics is compatible with the one of Lagrangian tracers in classical turbulence, while at short time scales the Magnus precession around the filaments caused by the vortex circulation is dominating the motion. Such information can be extracted consistently both in the frequency spectrum of the velocity and in the decay time of the correlation of the acceleration. Furthermore, if particles are light enough, faster frequencies are also excited. This suggests (as intuitively expected) that light particles can be more sensitive to the small scale fluctuations of the flow. 

Finally we investigated the particle velocity statistics. The distribution of the particles velocity is Gaussian, in contrast with the power law scaling $|v_i|^{-3}$ recently observed in superfluid helium experiments \cite{PaolettiVelStat2008,LaMantiaVelocity}. There are several reasons explaining why power-law tails are absent in our simulations. Firstly, since the simulation of each particle has an important numerical cost, the number of particles is restricted only to a couple of hundreds. Due to this issue, vortex reconnections might be unlikely sampled by the sparse distribution of particles. Note also that, as particles have a finite size, increasing their number keeping the size of the system constant will increase substantially the filling fraction.  In this case, turbulence could be even prevented by the presence of particles. Although interesting, this limit is out of the scope of this work. Secondly, the GP-model is compressible and particle moving at large velocities are slowed down by vortex nucleations. This certainly reduces large velocity fluctuations, perhaps limiting the development of power-law tails. It would be interesting to address such issues in generalized GP models, including a roton minimum and high-order non-linearities. Moreover, our simulations are by definition at zero temperature and particles do not follow the singular superfluid velocity field because of the lack of viscosity in the system. Indeed, in the GP model the pressure gradients that drive the particle dynamics are always regular because of the vanishing density at the vortex cores, unlike other models as the vortex filament method. As a consequence, the divergence of the superfluid velocity along the vortex lines can not be experienced by the particles. Conversely, at finite temperature the superfluid and the normal component can be locked thanks to mutual friction. In this case, since particles would sample the normal fluid velocity because of Stokes drag, they might be able to sample the $1/r$ flow around a quantum vortex. Finally, we observed that fast velocity fluctuations are highly sensitive to interpolation and filtering methods that could even lead to power-law tails. These tails are completely spurious, and special care is needed while analyzing numerical or experimental data.

\appendix
\section{Numerical artifacts on the particle velocity statistics. Comparison between B--spline and spectral interpolation methods}
\label{AppComp}
As explained in the main text, we evaluate the force $\mathbf{f}_i^\mathrm{GP}=-\left(V_\mathrm{p}*\nabla\rho\right)\left[\mathbf{q}_i\right]$ (\ref{Eq:GPPparticles}) at the particle position $\mathbf{q}_i$ using a B--spline interpolation method \cite{ToschiBspline} at each time step. Such method is precise and computationally cheap, but it turns out to present some issues that we have to take care of. In order to check the reliability of the method, we re-run a simulation using Fourier interpolation for one species of particles in the time window corresponding to the turbulent regime. Fourier interpolation is exact in the sense that uses the information of the full three dimensional field, that is resolved with spectral accuracy (i.e. discretization errors are at most exponentially small with the number of discretization points). The numerical cost of this method is the one of one Fourier transform (per particle). In Fig.\ref{Fig:CompSpeck} the velocity and acceleration spectra computed using B--spline and Fourier interpolation methods are compared. 
\begin{figure}[h!]
\includegraphics[width=.99\linewidth]{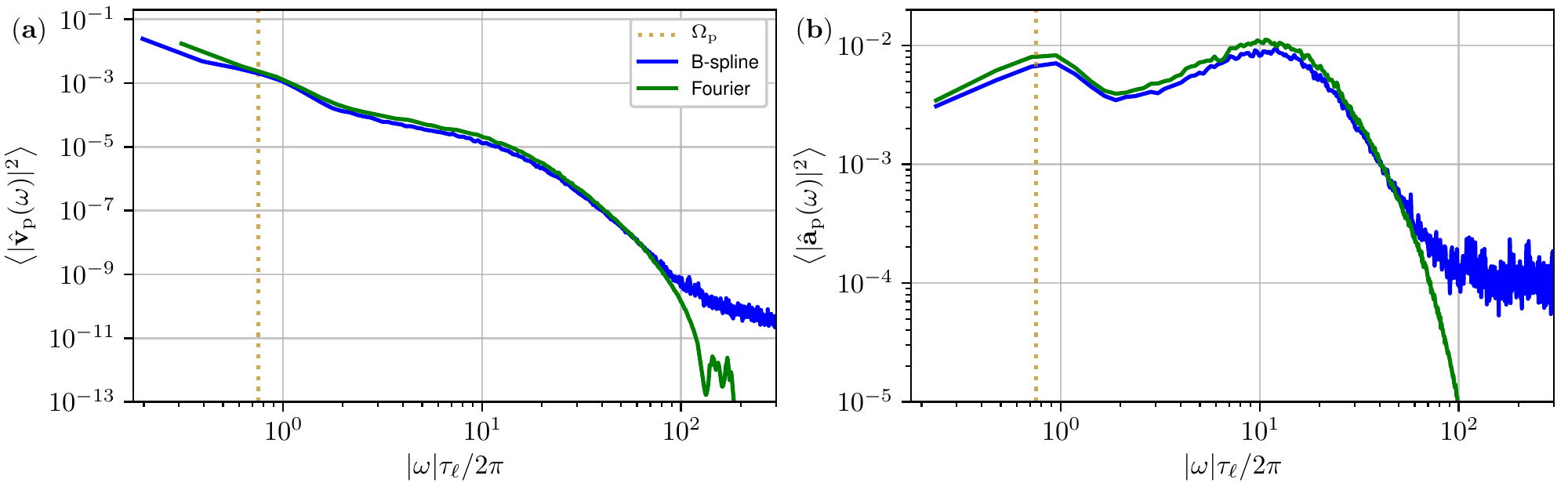}
\caption{(color online) Velocity spectra \textbf{(a)} and the acceleration spectra \textbf{(b)} for particles of size $a_\mathrm{p}=10\xi$ and mass $\mathcal{M}=0.13$, evolved using B--spline interpolation (blue lines) and spectral Fourier interpolation (green lines). The spectra are averaged over particles and over the times $1.3T_L<t<1.5T_L$. }
\label{Fig:CompSpeck}
\end{figure}
Clearly, the B--spline interpolation introduces non-physical fast oscillations, but at the frequencies $\omega<\omega_\mathrm{noise}=50(2\pi/\tau_\ell)$ the behavior of the spectra is unchanged. 
Nevertheless, some differences in the features of particle statistics are still visible at fast timescales once the noise is filtered out. 

We use a Gaussian convolution to perform a filtering of the velocity time series for each particle in the frequency window  $\omega_c<\omega<\omega_\mathrm{noise}$, where $\omega_\mathrm{c}$ is a variable infrared cut-off frequency. Then we compute the PDF of the filtered velocity for different values of $\omega_c$. Such PDFs are shown in Fig.\ref{Fig:CompVel} comparing the simulations in which Fourier and B--spline interpolation are used for the same species of particle. Surprisingly, only in the latter case we observe power-law tails for the fast oscillations distributions. Such PDFs are similar to the ones observed experimentally \cite{PaolettiVelStat2008,LaMantiaVelocity}, but in the present case, they are just a consequence of numerical artifacts. 
\begin{figure}[h]
\includegraphics[width=.99\linewidth]{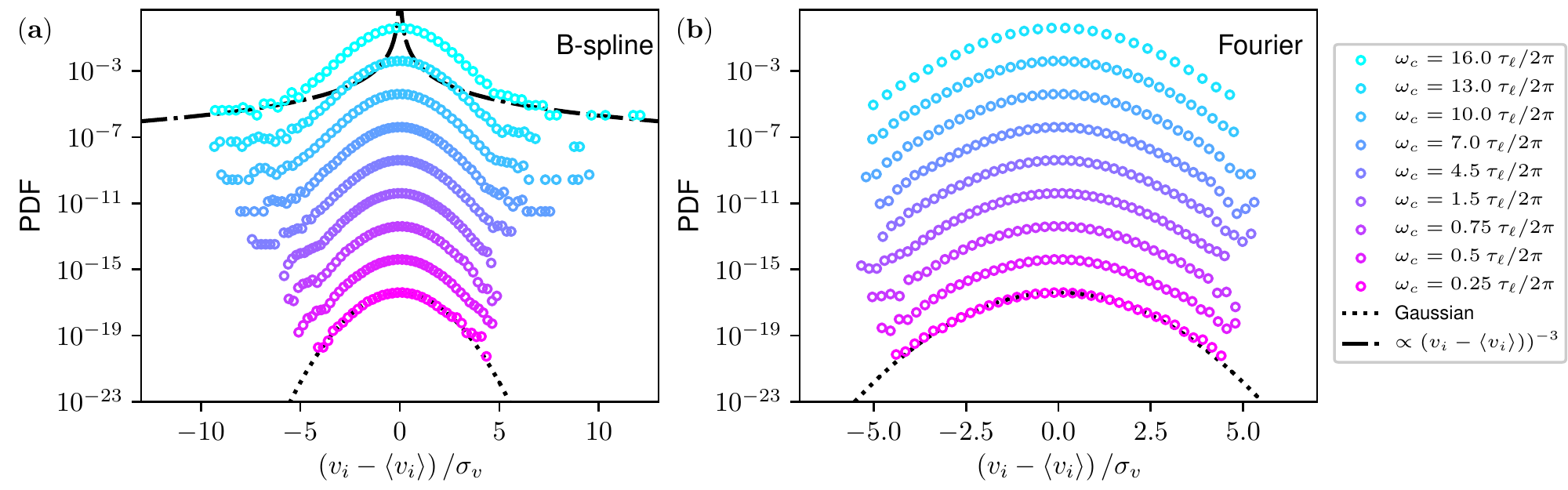}
\caption{(color online) Probability density function of the velocity filtered in the frequency window $\omega_c<\omega<\omega_\mathrm{noise}$ for different values of $\omega_c$.
Data refer to particles of size $a_\mathrm{p}=10\xi$ and mass $\mathcal{M}=0.13$. Dotted line is a gaussian distribution and dash-dotted line is a power-law scaling $0.002\left(v_i-\left\langle v_i \right\rangle\right)^{-3}$. The data are averaged over particles and over the times $1.3T_L<t<1.5T_L$. Different PDFs are shifted for visualization. \textbf{(a)} Particle force interpolated with B--spline method. \textbf{(b)} Particle force interpolated with Fourier method.}
\label{Fig:CompVel}
\end{figure}  

\acknowledgments
U.G. acknowledges J.I. Polanco for fruitful discussions. The authors were supported by Agence Nationale de la Recherche through the project GIANTE ANR-18-CE30-0020-01. Computations were carried out on the M\'esocentre SIGAMM hosted at the Observatoire de la C\^ote d'Azur and the French HPC Cluster OCCIGEN through the GENCI allocation A0042A10385.

%

\end{document}